\documentclass[aps,prd,reprint,superscriptaddress,twocolumn,preprintnumbers,floatfix,nofootinbib,amsmath,amssymb]{revtex4-2}

\usepackage{graphicx}
\usepackage{epstopdf}
\usepackage{mathrsfs}
\usepackage{amssymb}
\usepackage{verbatim}
\usepackage{xcolor}
\usepackage[dvipsnames]{xcolor}
\usepackage{multirow}
\usepackage{amsmath}
\usepackage{subfig}
 \usepackage{slashed}
\usepackage{float}
\usepackage[normalem]{ulem}
\usepackage{sidecap}
\usepackage{braket}
\usepackage{hyperref}
\usepackage{orcidlink}

\hypersetup{pdfstartview=FitV,colorlinks=true,linkcolor=blue,citecolor=red,filecolor=black,urlcolor=blue}

\usepackage{diagbox}

\usepackage[T1]{fontenc}
\usepackage[utf8]{inputenc}

\newcommand{\beq}{\begin{equation}}
\newcommand{\eeq}{\end{equation}}
\newcommand{\bea}{\begin{eqnarray}}
\newcommand{\eea}{\end{eqnarray}}

\def\to{\rightarrow}

\usepackage[parfill]{parskip}

\begin{document}

\vspace*{1mm}

\title{Addressing the Hubble tension with Sterile Neutrino Dark Matter}

\author{Debtosh Chowdhury\,\orcidlink{0000-0002-4302-7356}}
\email{debtoshc@iitk.ac.in}
\affiliation{Department of Physics, Indian Institute of Technology Kanpur, Kanpur 208016, India}

\author{Md~Sariful Islam\,\orcidlink{0009-0001-7174-7369}}
\email{sariful21@iitk.ac.in}
\affiliation{Department of Physics, Indian Institute of Technology Kanpur, Kanpur 208016, India}

\begin{abstract} 
\noindent One of the promising dark matter (DM) candidates is a keV-scale sterile neutrino. In the early universe, the observed relic of the sterile neutrino DM is generated via the \textit{Dodelson-Widrow} mechanism. However, this production scenario is severely constrained by various astrophysical observations. Many non-standard interactions between active ($\nu_a$) and sterile ($\nu_s$) neutrinos have been proposed to evade these astrophysical bounds. Here, we study sterile neutrinos in the context of a mass-varying scenario by coupling both active and sterile neutrinos to a scalar field. This novel mechanism opens up a new parameter space that generates the observed DM relic and  addresses the \textit{Hubble tension}. We find that the resulting parameter space can be fully probed by future X-ray missions. 
\end{abstract}

\date{\today}

\maketitle

\setcounter{equation}{0}

\section{Introduction}
Understanding the fundamental nature of dark matter (DM) remains one of the central open problems in astro-particle physics. Among the proposed candidates, sterile neutrinos with masses in the keV range have received significant attention \cite{Dodelson1994, Shi:1998km, Dolgov:2000ew, Abazajian:2001nj, Abazajian:2001vt, Asaka:2005an, Asaka:2005pn, Shaposhnikov:2006xi, Kusenko:2006rh, Petraki:2007gq, Petraki:2008ef, Loewenstein:2008yi, Loewenstein:2009cm, Drewes:2016upu, Merle:2017jfn, Goertz:2024gzw, Cline:2026cek, Koutroulis:2023fgp}. In the simplest scenario, sterile neutrinos are produced in the early Universe through oscillations with active neutrinos prior to their decoupling and before the onset of Big Bang Nucleosynthesis (BBN). This non-resonant production process is commonly referred to as the \textit{Dodelson-Widrow} (DW) mechanism \cite{Dodelson1994}. Sterile neutrinos are unstable and can undergo radiative decay into an active neutrino accompanied by a photon \cite{Pal:1981rm}. Such decays would generate monochromatic X-ray or gamma-ray lines, depending on the sterile neutrino mass, from regions where dark matter is concentrated, such as galaxies and galaxy clusters. However, results from dedicated X-ray and gamma-ray line searches place strong constraints on this scenario, creating significant tension with the DW production mechanism and excluding most of its viable parameter space \cite{Hofmann:2016urz, Borriello:2011un, Krivonos:2024yvm, Yin:2025xad}.

To evade the experimental bound, thus opening up new viable parameter space, many non-standard interactions (NSIs) have been proposed in the DW framework. These NSIs can be between active-active ($\nu_a - \nu_a$) neutrino \cite{DeGouvea:2019wpf, Kelly:2020pcy, Chichiri:2022} or active-sterile ($\nu_a - \nu_s$) neutrino \cite{Dev:2025}, or sterile-sterile ($\nu_s-\nu_s$) \cite{Astros:2023xhe, Bringmann:2022aim} neutrino sector. In the presence of an NSI, new scattering or annihilation channels open up in the active ($\nu_a$) and/or sterile ($\nu_s$) neutrino sector. So, the efficient production of the sterile neutrino from the active neutrino happens even for a small mixing angle.

Another unsolved problem in modern cosmology is \textit{Hubble tension} (see, e.g., \cite{DiValentino2021, Khalife2024, Vagnozzi:2023} for reviews). The tension arises from the persistent disagreement between determinations of the present-day expansion rate of the universe, $H_0$, inferred from early-universe observations and those obtained from late-universe measurements. Planck satellite measurement of cosmic microwave background (CMB) reported $H_0 = 67.4 \pm  0.5 \,\text{km}\,\text{s}^{-1}\text{Mpc}^{-1}$ \cite{Planck:2018vyg}. In contrast, local distance-ladder measurements based on Cepheid-calibrated Type Ia supernovae from the SH0ES collaboration report a significantly higher value, $ H_0$ = $73.30 \pm  1.04 \, \text{km}\,\text{s}^{-1}\,\text{Mpc}^{-1}$ \cite{Riess:2016jrr, Riess:2021jrx}.
So, there is a 5-$\sigma$ difference between the measured value of $H_0$ by Planck and SH0ES. To alleviate this tension, many modifications to the $\Lambda$CDM cosmology have been proposed \cite{SmithPoulinAmin2020, BeneventoHuRaveri2020, CliftonHyatt2024, Kouniatalis:2026avj, Dainotti:2025qxz}. One such solution is early dark energy (EDE) \cite{Kamionkowski:2022pkx, KarwalKamionkowski2016, SaksteinTrodden2020, PoulinSmithKarwalKamionkowski2019, Rezazadeh:2024}. Refs.\cite{PhysRevD.111.063551, SaksteinTrodden2020} studied EDE by quadratically coupling the Standard Model (SM) neutrinos to a background scalar field. Because of this coupling, the ensemble of SM neutrinos, i.e., active neutrinos, creates an effective potential for the scalar, but the vacuum expectation value (vev) of the scalar back-reacts and gives active neutrinos an effective mass. These effects generate the required dynamics and can address the \textit{Hubble tension}. This tension can also be eased by increasing $\Delta N_{\text{eff}}$ \cite{Vagnozzi:2020} via the decay of a sterile neutrino of mass $\mathcal{O}(30)$ MeV \cite{GelminiKusenkoTakhistov2021} or $\mathcal{O}(150-450)$ MeV \cite{GelminiKawasakiKusenkoMuraiTakhistov2020,Dolgov:2000jw} at BBN.

In this study, we assume that the background scalar field is coupled to both the sterile and active neutrinos. Hence, the sterile neutrino also gets an effective mass due to the vev of the background scalar field. This feature is unique compared to other NSI or lepton-asymmetry models \cite{Shi:1998km} for sterile neutrinos. In those models, the mass of the sterile neutrino remains the same throughout its cosmological history, from production until today. Whereas, in our case, the sterile neutrino mass varies as the background scalar field evolves over time. We have shown that, due to an increased effective sterile neutrino mass in the early universe, a viable parameter space in the mixing-angle--mass plane opens up, evading existing experimental constraints. We further show that in the allowed region of the sterile neutrino parameter space, the inferred value of $H_0$ is modified compared to the standard $\Lambda$CDM scenario and can reproduce the SH0ES-measured value.

We organize the rest of the paper as follows. We describe our model in Sec.~II and discuss the cosmological evolution of the scalar field and the sterile neutrino in Sec.~III. The result is presented in Sec.~IV, and we conclude in Sec.~V.

\section{Model Setup}
In this model, an active neutrino and a sterile neutrino are quadratically coupled with a real homogeneous scalar field $\phi$. The interaction Lagrangian is given by 
\begin{equation}
    -\mathcal{L}_{\mathrm{int}} \supset \epsilon\frac{m_{\nu_{1}}}{M_{\mathrm{pl}}^2}\bar{\nu}_1\nu_1\phi^2 + \epsilon\frac{m_{\nu_{4}}}{M_{\mathrm{pl}}^2}\bar{\nu}_4\nu_4\phi^2.
    \label{eq. Lagrangian}
\end{equation}
Here, $\epsilon$ is the coupling strength, $M_{\mathrm{pl}}$ is the Planck mass, $m_{\nu_1}$, and $m_{\nu_4}$ are Dirac masses for active and sterile neutrinos, respectively. As we impose a $\mathbb{Z}_2$ symmetry ($\phi \to - \phi$) on the Lagrangian, the above dimension-five operator is the leading-order interaction term between the neutrinos and the scalar field $\phi$ \footnote{We assume that the scalar field self-interactions are negligible. For the effect of the quartic self-interaction term of the scalar field on the Hubble tension, see Ref.~\cite{LopezSanchez2025}.} \cite{Sibiryakov:2020nzo}.
In the absence of any NSIs in the neutrino sector, sterile and active neutrino flavors mix through the vacuum mixing angle $\theta$.
So, we can express the sterile and active vacuum mass eigenstate in terms of their flavor eigenstate as
\begin{eqnarray}
    \ket{\nu_1} =&& \, \cos{\theta} \ket{\nu_{a}} - \sin{\theta} \ket{\nu_{s}} \, , \nonumber  \\
    \ket{\nu_4} =&& \, \cos{\theta} \ket{\nu_{s}} + \sin{\theta} \ket{\nu_{a}}.
    \label{eq. vacuum mixing}
\end{eqnarray}
Active and sterile neutrinos are produced on-shell during their cosmological evolution. Therefore, using the equation of motion, effective neutrino masses can be written as~\cite{PhysRevD.111.063551}
\begin{eqnarray}
    m_{\nu_{1}, \text{eff}} =&& \, m_{\nu_{1}} \left( 1+ \epsilon\frac{\phi^2}{M_{\mathrm{pl}}^2}\right) \, ,\nonumber \\
    m_{\nu_{4}, \text{eff}} =&& \, m_{\nu_{4}} \left( 1+ \epsilon\frac{\phi^2}{M_{\mathrm{pl}}^2}\right).
    \label{eq. effective mass}
\end{eqnarray}
As can be seen from Eq.~\eqref{eq. effective mass}, a universal coupling $\epsilon$ rescales both the active and sterile neutrino masses by the same factor, leaving the vacuum mixing angle $\theta$ unchanged. Motivated by this observation, and to keep the model as minimal as possible, we assume the coupling $\epsilon$ to be the same for both the active and sterile neutrino sectors.

The active neutrino remains in thermal equilibrium via the SM weak interactions before BBN.  Around 1 MeV the active neutrino decouples from the thermal plasma with a relativistic Fermi-Dirac distribution. As for the sterile neutrino, it will never come to thermal equilibrium in our scenario, but we can approximately model its phase space density using the following Fermi-Dirac form \cite{Dev:2025}
\begin{equation}
    \label{eq. sterile psd}
    f_s(p, T) = \frac{\alpha}{\exp{\left(p/T_s\right)}+1}\, ,
\end{equation}
where $T_s$ is the effective temperature, $p$ is the sterile neutrino momentum, and $\alpha$ represents the deviation from thermal equilibrium with value in the range $0< \alpha \lesssim 1$. In general, both $T_s$ and $\alpha$ are functions of the SM bath temperature $T$ \cite{Dev:2025, dodelson:1996}.

 In an isotropic and homogeneous Friedmann--Lemaître--Robertson--Walker (FLRW) background, the Euler--Lagrange equation corresponding to the interaction Lagrangian in Eq.~\eqref{eq. Lagrangian} yields the following equation of motion for the homogeneous scalar field
\begin{equation}
    \label{eq. evolution of phi}
    \frac{d^2\phi}{dt^2} + 3H(t)\frac{d\phi}{dt} + \left( m_{\phi}^2 + 2\epsilon\frac{m_{\nu_{1}}}{M_{\mathrm{pl}}^2}\bar{\nu}_1\nu_1 + 2\epsilon\frac{m_{\nu_{4}}}{M_{\mathrm{pl}}^2}\bar{\nu}_4\nu_4\right)\phi = 0,
\end{equation}
where $m_{\phi}$ is the bare mass of the scalar field $\phi$, $t$ is the proper time, and $ H(t)$ is the total Hubble rate including contributions from active and sterile neutrinos and the scalar field.
At the quantum level, the terms proportional to $\bar{\nu}_{1}\nu_{1}$ and $\bar{\nu}_{4}\nu_{4}$ are substituted by their corresponding expectation values evaluated in their respective neutrino backgrounds. This expectation value is given by (see, for example, Appendix A of Ref.~\cite{Smirnov:2022sfo})
\begin{equation}
    \label{eq. psibarpsivalue}
     \braket{\bar{\nu}_{i}\nu_{i}} = \int \frac{d^3{p}}{(2\pi)^3}
    \frac{m_{\nu_{i,\text{eff}}}}{E_{{p}}} f({p}) = \frac{\rho_{\nu_{i}}-3P_{\nu_{i}}}{m_{\nu_{i,\text{eff}}}},
\end{equation}
with $i = 1, 4$ corresponding to active and sterile neutrinos, respectively. Here, $\rho_{\nu_i}$ and $P_{\nu_i}$ are the energy density and pressure of the $i^{\rm th}$ neutrino species, respectively.
Then using Eq.~\eqref{eq. psibarpsivalue}, we can rewrite the effective potential in Eq.~\eqref{eq. evolution of phi} as \cite{SaksteinTrodden2020} 
\begin{align}
    \label{eq. effective potential}
      \frac{1}{\phi}\frac{\partial\,V_{\text{eff}}}{\partial\,\phi} = m_{\phi}^2 +
     2\epsilon\,  \frac{m_{\nu_{1}}}{M_{\mathrm{pl}}^2}\frac{\rho_{\nu_{1}}-3P_{\nu_{1}}}{m_{\nu_{1,\text{eff}}}} \\  \nonumber + 2\epsilon\, \frac{m_{\nu_{4}}}{M_{\mathrm{pl}}^2}\frac{\rho_{\nu_{4}}-3P_{\nu_{4}}}{m_{\nu_{4,\text{eff}}}}.
\end{align}
As a result, the pressure and energy density of both neutrinos affect the evolution of the background scalar field through their contribution to the effective potential $V_{\rm eff}$.

At any time $t$, the Hubble rate receives a contribution from the standard $ \Lambda\mathrm{CDM} $ part plus an extra contribution due to the changes of effective masses of $\nu_1$ and $\nu_4$ for their interaction with the background scalar field $\phi$, i.e.,  
\begin{eqnarray}
    \label{eq. modified hubble}
    H^2 = H_{\Lambda \mathrm{CDM}}^2 + \frac{8 \pi}{3M_{\mathrm{pl}}^2}\rho_{\mathrm{extra}},
\end{eqnarray}
where $ \rho_{\mathrm{extra}}$ is 
\begin{eqnarray}
    \label{eq. rho extra}
    \rho_{\mathrm{extra}} = &&\, \rho_{\nu_1}\left( m_{\nu_{1}, \text{eff}} , T_{\nu_{1}}\right) - \rho_{\nu_1}\left( m_{\nu_{1}} , T_{\nu_{1}}\right) + \\  \nonumber && \rho_{\nu_4}\left( m_{\nu_{4}, \text{eff}} \right)  - \rho_{\mathrm{CDM}} 
    + \frac{1}{2} \left(\frac{d\phi}{dt}\right)^2 + \frac{1}{2}m_{\phi}^2 \phi^2.
\end{eqnarray}
Here, $\rho_{\nu_1}\left( m_{\nu_{1}, \text{eff}}, T_{\nu_{1}}\right)$ and $\rho_{\nu_4}\left( m_{\nu_{4}, \text{eff}} \right)$  are the modified active and sterile neutrino energy density due to their effective masses, respectively. Also, $\rho_{\nu_1}\left( m_{\nu_{1}}, T_{\nu_{1}}\right)$ is the active neutrino energy density in $\Lambda\text{CDM}$ for $ m_{\nu_1} = 0.05\, \text{eV}$ and $\rho_{\mathrm{CDM}}$ is the cold dark matter energy density. As $\rho_{\nu_1}\left( m_{\nu_{1}} , T_{\nu_{1}}\right)$ and $\rho_{\mathrm{CDM}}$  are already included in $ H_{\Lambda \mathrm{CDM}}$, we subtracted those terms from Eq.~\eqref{eq. rho extra}.

The production of sterile neutrinos proceeds via oscillations between active and sterile neutrino states in the early universe, prior to Big Bang Nucleosynthesis (BBN), through the so-called Dodelson-Widrow mechanism \cite{Dodelson1994}. The evolution of the sterile neutrino phase-space distribution is described by a semi-classical Boltzmann equation of the form \cite{Abazajian:2001nj, Dolgov:2000ew, Stodolsky:1987, Rfoot:1997, DiBari:1999ha, Lee:2000}
\begin{widetext}
\begin{eqnarray}   
\label{eq. Boltzmann}
\frac{\partial}{\partial t} f_s\left(p,t\right) -H p \frac{\partial}{\partial p}f_s\left(p,t\right) \approx
    \dfrac{1}{4}\dfrac{\Gamma_{a}(p)\Delta^2(p)\sin^2{2\theta}}{\Delta^2(p)\sin^2{2\theta} + \left(\Gamma_{a}(p)\right)^2/4 + \left[ \Delta(p)\cos{2\theta}-V_{{T}}\right]^2}
    \times\left[ f_{a}\left(p,t\right)-f_{s}\left(p,t\right)\right],
\end{eqnarray}
\end{widetext}
where $f_s\left(p,t\right)$ and $f_a\left(p,t\right)$  denote the sterile neutrino momentum distribution function, and the active neutrino Fermi-Dirac distribution function, respectively. Here, $ \Gamma_{a}(p)$ is the active neutrino forward interaction rate and $\Delta(p) = \frac{m_{\nu_{4},\text{eff}}^2-m^2_{\nu_{1},\text{eff}}}{2p}$.  The effective matter-mixing angle is given by
\begin{equation}
\label{eq. matter angle}
    \sin^2{2\theta_m} = \frac{\Delta^2(p)\sin^2{2\theta}}{\Delta^2(p)\sin^2{2\theta}  + \left[ \Delta(p)\cos{2\theta}-V_{{T}}\right]^2}.
\end{equation}
In the SM, $\Gamma_{a}(p)$ is obtained by summing over all leptonic and hadronic channels, and it is expressed as \cite{Abazajian:2001nj, Asaka:2006nq, Merle:2016}
\begin{align}
\label{eq. interaction rate}
    \Gamma_{a}(p) = C_{a}(T) G_{\mathrm{F}}^2 \,p \,T^4,
\end{align}
where $ C_{a}(T)$ is a temperature-dependent function. We have numerically evaluated $\Gamma_{a}(p)$ from Ref.~\cite{PhysRevD.94.043515}. The SM thermal potential $ V_T$ is given by \cite{Notzold:1987ik, Abazajian:2006, DOlivo:2016}
\begin{equation}
    \label{eq. thermal potential}
    V_T(p) = - \dfrac{8\sqrt{2}G_{\mathrm{F}} p}{3m_W^2}\left[\left(\rho_{l_{a}} + \rho_{\Bar{l}_{a}}\right)
    + \dfrac{m_{W}^{2}}{m_Z^2}\left(\rho_{\nu_{a}} + \rho_{\Bar{\nu}_{a}}\right)\right],
\end{equation}
where $\rho_{l_{a}} (\rho_{\Bar{l}_{a}})$ is the energy density of the $a^{\rm th}$ flavor charged lepton ($l_{a}$) (anti-lepton ($ \Bar{l}_{a}$)), $\rho_{\nu_{a}} (\rho_{\Bar{\nu}_{a}})$ is the energy density of the $a^{\rm th}$ flavor active neutrino, and $p$ is the active neutrino momentum. As the NSI terms are suppressed by $M_{\mathrm{pl}}^2$, the scattering channel $\nu_{a}\nu_{s} \leftrightarrow \nu_{a}\nu_{s}$ or annihilation channel  $\nu_{a}\nu_{a} \leftrightarrow \nu_{s}\nu_{s}$ will be highly suppressed compared to the SM weak interaction channels (i.e., $\nu_a + \nu_a \leftrightarrow \nu_a + \nu_a, l_a + \nu_a \leftrightarrow l_a + \nu_a$ etc). Hence, the contribution from the NSI terms in $ \Gamma_{a}(p)$ or $ V_{T}$ is strongly suppressed.

\section{Cosmological evolution of the sterile neutrino}
\begin{figure}
    \centering
    \includegraphics[width=0.9\linewidth]{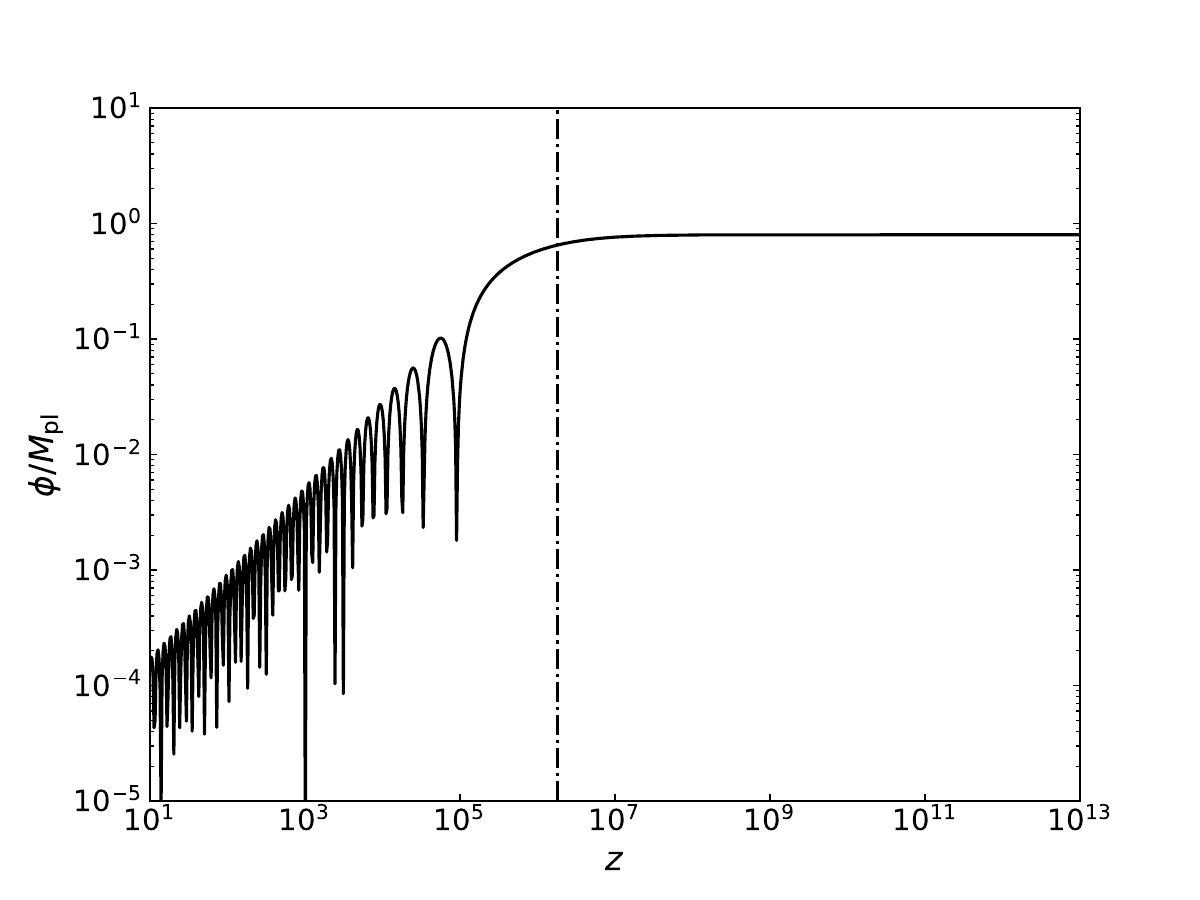}
    \caption{\small Evolution of the scalar field with respect  to the redshift \( z\) for $\phi_0/M_{\text{pl}}$ = 0.8 and coupling, $\epsilon = 1.56\times 10^3$. Here, the sterile neutrino mass is set to 10 keV, and the effective mass is 10 MeV. The vertical dash-dotted line corresponds to $z_{\text{osc}} = 1.81\times 10^6$.}
    \label{fig. phi field oscillation}
\end{figure}

\begin{figure}
    \centering
    \includegraphics[width=\linewidth]{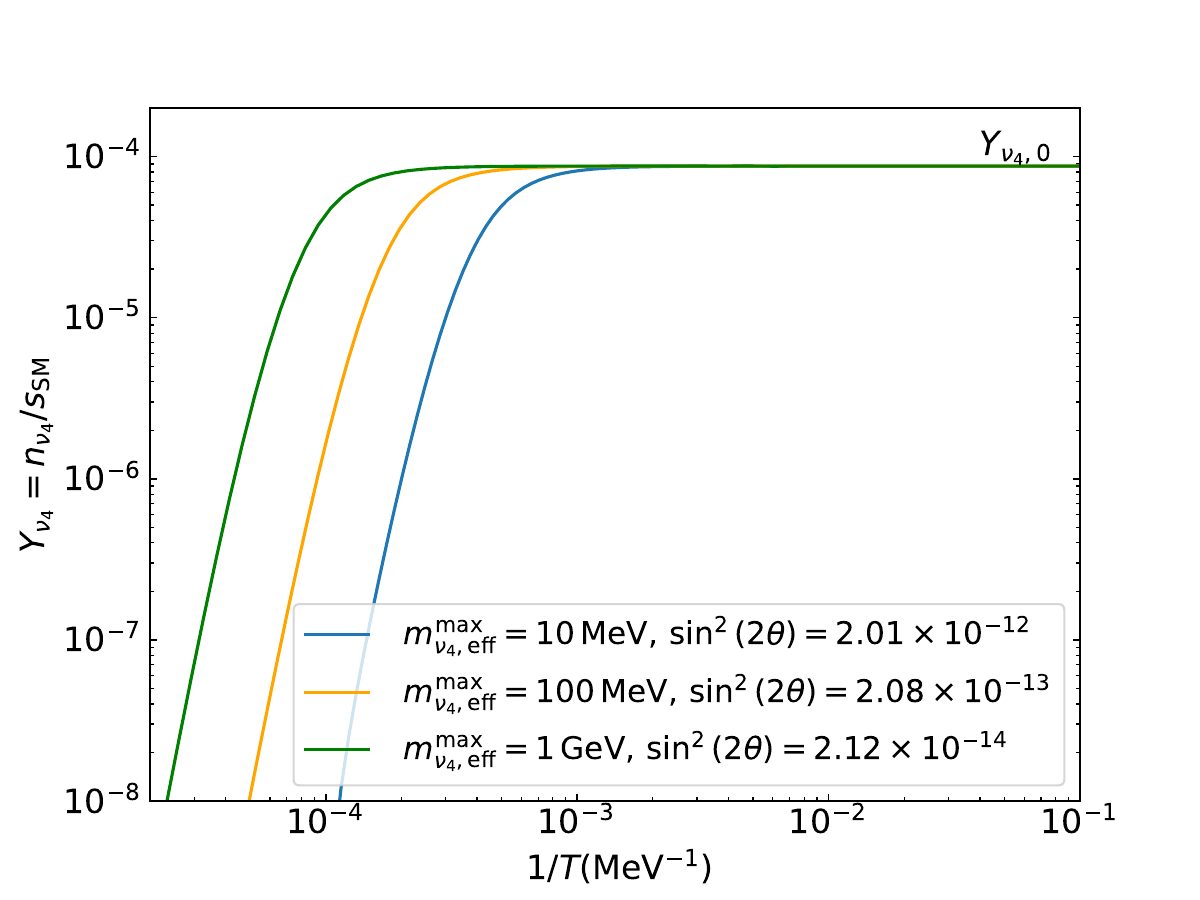}
    \caption{\small Yield of sterile neutrino for present-day sterile mass of $ m_{\nu_{4}} = 10 \, \text{keV}$. Increasing  $m^{\text{max}}_{\nu_{4},\, \text{eff}}$ by one order leads to production at a higher temperature and reduces the vacuum mixing angle by one order. }
    \label{fig: sterile neutrino yield}
\end{figure}

In the early universe at some high temperature ($T \gtrsim 100\,\text{GeV}$), the scalar field $\phi$ is displaced from the minimum of its potential to some value $\phi_0$, close to the Planck scale. This displacement of the scalar field will come from some UV dynamics. Since during this period $H^2 \gg \frac{1}{\phi}\frac{\partial\,V_{\text{eff}}}{\partial\,\phi}$, the evolution of $\phi$, governed by Eq.\eqref{eq. evolution of phi}, is highly overdamped leaving the scalar field frozen at $\phi_0$ (see Fig.~\ref{fig. phi field oscillation}).
So, the effective mass of the sterile neutrino is constant initially and is given by $m^{\text{max}}_{\nu_{4}, \text{eff}} = \, m_{\nu_{4}} \left( 1+ \epsilon\frac{\phi_0^2}{M_{\mathrm{pl}}^2}\right)$. For a given $\epsilon$ and $\phi_0$, $m^{\text{max}}_{\nu_{4}, \text{eff}}$ is the maximum effective mass of the sterile neutrino. With this enhanced effective mass, the sterile neutrino ($\nu_{4}$) is produced from the active neutrino ($\nu_{1}$) via the DW mechanism.  We calculate the number density of the sterile neutrino ($n_{\nu_4}$) by numerically solving Eq.~\eqref{eq. Boltzmann} after modifying the publicly available \texttt{sterile-dm} package \cite{PhysRevD.94.043515}.
In Fig.~\ref{fig: sterile neutrino yield}, we show the evolution of the sterile neutrino yield, $Y_{\nu_4} \equiv n_{\nu_{4}}/s_{\rm SM}$, where $s_{\rm SM}$ is the entropy density of the background radiation, for temperatures ranging from 50 GeV to 10 MeV for present-day sterile neutrino mass $m_{\nu_4} = 10$ keV. Blue, orange, and green contours correspond to early-universe sterile neutrino masses, $m^{\text{max}}_{\nu_{4},\, \text{eff}}$ = 10 MeV, 100 MeV, and 1 GeV, respectively.
From Fig.~\ref{fig: sterile neutrino yield} we observe that increasing $m^{\text{max}}_{\nu_{4},\, \text{eff}}$ triggers the production of the sterile neutrino at increasingly high temperatures.  This is due to the fact that $\Delta(p)$ is proportional to the effective sterile neutrino mass $(m_{\nu_{4}, \text{eff}})$. As a result, increasing $m_{\nu_{4}, \text{eff}}$ extends the temperature interval over which  $\Delta(p)$ dominates over the thermal potential ($V_T$), relative to the standard DW scenario. For a fixed $m_{\nu_{4}}$, this yields a larger effective matter-mixing angle ($\theta_{m}$), see Eq.\eqref{eq. matter angle}, compared to its value in the standard DW scenario. This effect leads to an enhanced production of sterile neutrinos at high temperatures. The sterile neutrino production reaches a maximum at temperature \cite{Dodelson1994} 
\begin{eqnarray}
\label{eq. Tmax}
   T_{\text{max}} \approx 133\, \text{MeV} \left(\frac{m^{\text{max}}_{\nu_{4},\, \text{eff}}}{\text{1\,keV}}\right)^{1/3}.
\end{eqnarray} 
For $T \gg T_{\text{max}}$, thermal potential $V_T$ dominates over $\Delta(p)$. Hence, using Eq.\eqref{eq. Boltzmann} we deduce that $Y_{\nu_4} \propto \left(m^{\text{max}}_{\nu_{4}, \,\text{eff}}\right)^{4}\sin^2(2\theta)/T^9$. At $ T \ll T_{\text{max}}$, sterile co-moving number density freezes out and the yield $ Y_{\nu_4}$ becomes constant until today. The freeze-out occurs before the BBN epoch in our parameter space. The sterile neutrino relic density can then be computed using the relation \( \Omega_{\text{DM}} = Y_{\nu_4,0}\,s_0\,m_{\nu_4}/\rho_0\), where $Y_{\nu_4,0}$, $ s_0$, and $ \rho_0$ are the sterile neutrino yield, entropy density, and critical energy density today, respectively. Utilizing Eq.\eqref{eq. Tmax}, the sterile neutrino relic density is 
\begin{equation}
    \label{ eq. relic}
    \Omega_{\text{DM}} \propto \left(\frac{m^{\text{max}}_{\nu_{4},\text{eff}}}{\text{keV}}\right)\left(\frac{m_{\nu_4}}{\text{keV}}\right)\sin^2(2\theta).
\end{equation}
Therefore, to satisfy the observed DM relic density, $\omega_{\text{DM}} = \Omega_{\text{DM}} h^2 = 0.12$ \cite{Planck:2018vyg}, we can accommodate a smaller mixing between the active and sterile neutrino, than in the standard DW scenario,  by increasing $m^{\text{max}}_{\nu_{4},\, \text{eff}}$.  From Fig.~\ref{fig: sterile neutrino yield}, we see that for a fixed $m_{\nu_4}$, a change in $ m^{\text{max}}_{\nu_{4}, \text{eff}}$  can be compensated by inversely tuning $ \sin^2{2\theta}$.

The $\phi$ starts oscillating when the effective potential of the scalar field dominates over the Hubble friction term i.e., \( \frac{1}{\phi}\frac{\partial\,V_{\text{eff}}}{\partial\,\phi} \gg H^2\). As the active neutrino ($\nu_{1}$) is relativistic during the entire period that is relevant to our analysis, it behaves as radiation-like, i.e., \( P_{\nu_1} \approx \frac{1}{3}\rho_{\nu_1}\). The mass of the scalar field should be $m_{\phi} \lesssim H( z_{\text{CMB}}) \approx 10^{-29}\,\text{eV}$ such that the $\phi$ decays by the time of CMB and it has a negligible contribution to the Hubble parameter. Therefore, we set $ m_{\phi} = 0$ in our numerical analysis. So, from Eq.\eqref{eq. effective potential} we see that \( \frac{\partial\,V_{\text{eff}}}{\partial\,\phi}\) is dominated by the contribution from the sterile neutrino ($\nu_{4}$).
The sterile neutrino becomes non-relativistic near the BBN epoch in our region of interest. Hence, ignoring the pressure term for the sterile neutrino in Eq.\eqref{eq. effective potential}, the redshift at which the scalar field starts oscillating,  determined by equating \( \frac{1}{\phi}\frac{\partial\,V_{\text{eff}}}{\partial\,\phi} = H^2\), is given by
\begin{equation}
\label{eq. oscillation redshift}
    z_{\text{osc}} \simeq \frac{3\,\epsilon}{4\, \pi} \frac{\omega_{\text{DM}}}{\omega_{\text{rad}}},
\end{equation}
where $\omega_{\text{DM}}$ and $ \omega_{\text{rad}}$ are the present-day relic density of DM and radiation, respectively.  In deriving Eq.\eqref{eq. oscillation redshift} we have assumed oscillation starts during the radiation-dominated (RD) epoch. From Fig.~\ref{fig. phi field oscillation} we see that the value of $z_{\text{osc}}$, represented by the vertical dash-dotted line, is of the order of $10^6$. Amplitude of the oscillation of $ \phi$, calculated using the WKB approximation \cite{Sibiryakov:2020nzo}, is
\begin{equation}
\label{eq. phi amp}
    \phi_{\text{amp}} \propto \left(\frac{1}{\phi}\frac{\partial\,V_{\text{eff}}}{\partial\,\phi}\right)^{-1/4}\,(1+z)^{3/2} \propto (1+z)^{3/4}.
\end{equation}
Hence, the kinetic energy density of the scalar field redshifts away as \( \frac{1}{2} \left(\frac{d\phi}{dt}\right)^2 \propto (1+z)^{11/2}\) in RD and as \( \frac{1}{2} \left(\frac{d\phi}{dt}\right)^2 \propto (1+z)^{9/2}\) during MD which is faster than any other components of the universe. So, the scalar field will not have any relic in the present-day.

The non-trivial evolution of $\rho_{\nu_4}$ happens during the oscillation of the $\phi$ field. From Eqs.~\eqref{eq. effective mass} and \eqref{eq. phi amp}, the effective sterile neutrino mass and hence energy density redshift as \( \rho_{\nu_4} \propto (1 +z)^{9/2}\) during this epoch. So, we can summarize the evolution of $ \rho_{\nu_4}$ as follows
\begin{equation}
    \label{eq. evolution of rho4}
    \rho_{\nu_4} \propto \left\{
    \begin{array}{ll}
      m_{\nu_4} \,\epsilon\frac{\phi_0^2}{M_{\text{pl}}^2}(1+z)^3,   & \qquad z > z_{\text{osc}}  \\[4pt]
        m_{\nu_4} \,\epsilon\frac{\phi_0^2}{M_{\text{pl}}^2}(1+z)^{9/2}, & \qquad z_{\text{decay}} < z < z_{\text{osc}} \\[4pt]
        m_{\nu_4}\,(1+z)^3, & \qquad z< z_{\text{decay}}
    \end{array}
    \right.
\end{equation}
where $ z_{\text{decay}} = \left( 10^{-2}\times \frac{M_{\text{pl}}^2}{\epsilon\,\phi_0^2}\right)^{2/3}\times z_{\text{osc}}$, is the redshift value when effective sterile neutrino mass returns to approximately 1\% of its present-day mass.
We see from Eq.\eqref{eq. evolution of rho4} that in our scenario, before the oscillation of the $\phi$ field, the sterile neutrino energy density is greater than the cold dark matter density in the $\Lambda \text{CDM}$ scenario by a factor of $\epsilon\frac{\phi_0^2}{M_{\text{pl}}^2} $. So, in our scenario, in the early universe, the total energy density, hence the Hubble parameter value, will be larger in comparison to the Hubble parameter value in the $\Lambda \text{CDM}$ scenario.
\begin{figure}
    \centering
    \includegraphics[width= 0.9\linewidth]{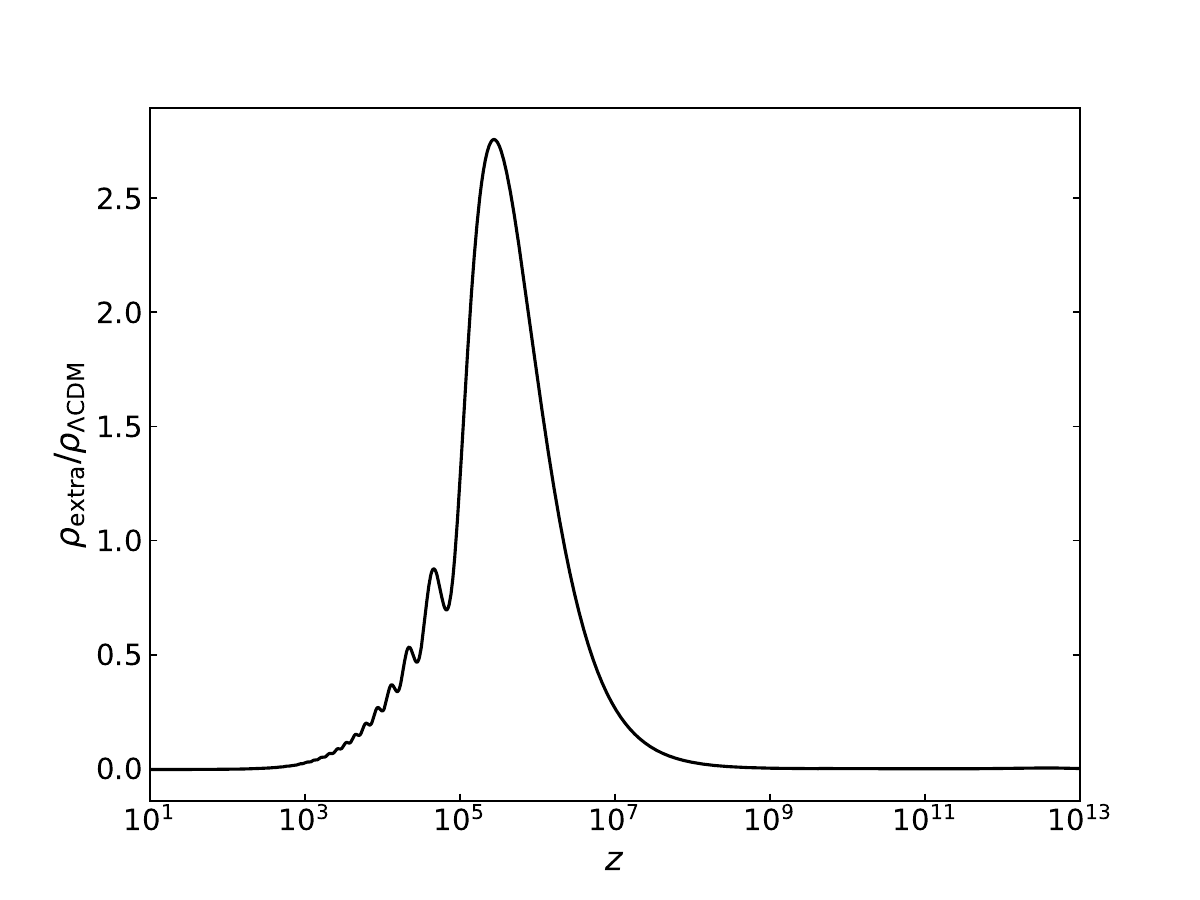 }
    \caption{\small Evolution of extra energy density ($\rho_{\text{extra}}$) with respect to the redshift ($z$) for $\phi_0/M_{\text{pl}}$ = 0.8 and coupling, $\epsilon = 1.56\times 10^3$. Here, the present-day sterile neutrino mass is 10 keV, and the effective mass is 10 MeV. $\rho_{\text{extra}}$ has a contributuion of the order of $\rho_{\Lambda\text{CDM}}$ in the redshift range $z \approx 10^7-10^3$.  }
    \label{fig: extra rho}
\end{figure}
In Fig.~\ref{fig: extra rho}, we depict the evolution of this extra contribution to the energy density as a function of redshift. Between the redshift $10^7-10^3$, $\rho_{\text{extra}}$ has a significant contribution of the order of $ \rho_{\Lambda \text{CDM}}$. This extra energy density before photon decoupling changes the sound horizon at the surface of last scattering. This, in turn, alters the inferred value of $H_0$. The change in $H_0$ can be estimated as \cite{Kamionkowski:2022pkx, PhysRevD.111.063551} 
\begin{eqnarray}
    \label{eq. H0}
    1+ \frac{\delta H_0}{H_0} &&\approx \frac{\int_{z_\text{CMB}}^{\infty} dz \left( \rho_{\Lambda \text{CDM}}(1+R(z))\right)^{-1/2}}{\int_{z_\text{CMB}}^{\infty} dz \left( (\rho_{\Lambda \text{CDM}}+\rho_{\text{extra}})(1+R(z))\right)^{-1/2}}\nonumber\\
   && \times  \frac{\int^{z_\text{CMB}}_{0} dz \left( \rho_{\Lambda \text{CDM}} + \rho_{\text{extra}}\right)^{-1/2}}{\int^{z_\text{CMB}}_{0} dz \left( \rho_{\Lambda \text{CDM}}\right)^{-1/2}},
\end{eqnarray}
where $ R(z) = \frac{3}{4}\frac{\omega_{b}}{\omega_{\gamma}}(1+z)^{-1}$ and $\omega_b = 2.24\times 10^{-2}$, $ \omega_{\gamma} = 2.47\times 10^{-5}$ are the present-day baryon and photon relic densities in $\Lambda \text{CDM}$, respectively. We have taken the redshift of recombination to be $z_{\text{CMB}} = 1080$ and the central value of $H_{0}$ in $\Lambda \text{CDM}$ to be $ 67.4\, \text{km}\,\text{s}^{-1}\,\text{Mpc}^{-1}$.

\section{Results}
\begin{figure}
    \centering
    \includegraphics[width=0.9\linewidth]{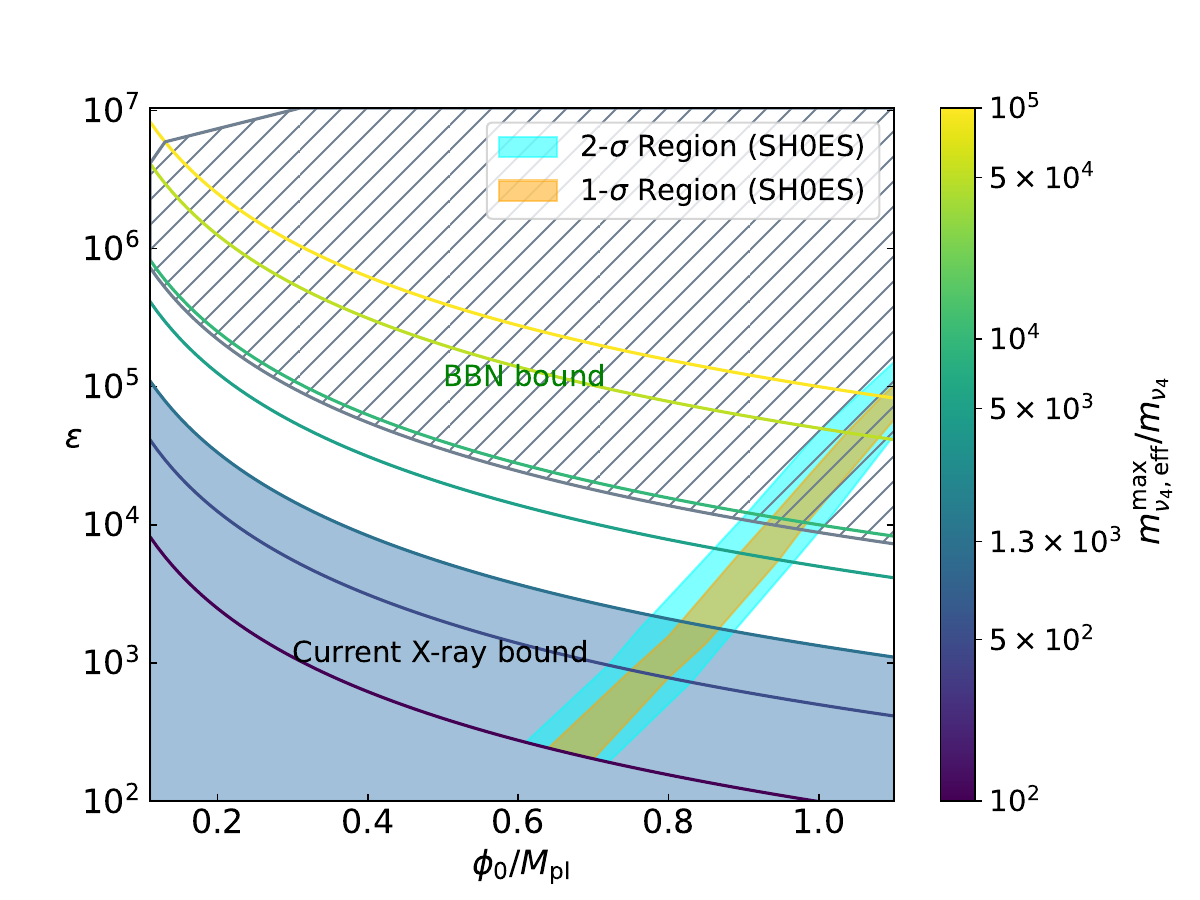}
    \caption{\small Scalar field parameter space for $m_{\nu_4}$ = 10 keV.  Each contour line corresponds to different $m^{\text{max}}_{\nu_4, \text{eff}}/m_{\nu_4}$ and satisfies the DM relic, $\Omega_{\text{DM}} h^2 = 0.12$, for specific values of $\sin^2{2\theta}$. The $H_0$ value increases as we move from left to right along each contour. We have marked the SH0ES 1-$\sigma$ and 2-$\sigma$ regions in orange and cyan, respectively. The region below $m^{\text{max}}_{\nu_4,\text{eff}} / m_{\nu_4} = 1.3 \times 10^3$ (blue-shaded area) is excluded by X-ray observations. The hatched region above $m^{\text{max}}_{\nu_4,\text{eff}} / m_{\nu_4} = 8.8 \times 10^3 $ is excluded by BBN constraints.}
    \label{fig:scalarparspace}
\end{figure}
In Fig.~\ref{fig:scalarparspace}, we delineate the parameter space for the scalar field $\phi$ that reproduces the SH0ES measurement of the Hubble parameter. In each contour line, $m^{\text{max}}_{\nu_4, \text{eff}} / m_{\nu_4}$ is constant with $m_{\nu_4} = 10$ keV and satisfies the DM relic for a specific value of $\sin^{2}2\theta$.  On each contour, the mixing angle ($\theta$) required to satisfy the DM relic density is smaller as we increase the ratio of $m^{\text{max}}_{\nu_4, \text{eff}} / m_{\nu_4}$. For example, increasing $m^{\text{max}}_{\nu_4 , \text{eff}} / m_{\nu_4}$ from $10^3$ to $10^4$  requires reducing the $\sin^{2}2\theta$ from $2.01\times10^{-12}$ to $2.08\times10^{-13}$, in order to satisfy relic density. As a result, in our scenario, for a fixed sterile neutrino mass, the X-ray bounds on the sterile neutrino can be evaded as we require a smaller mixing angle compared to the standard DW case.

Along each contour, moving from left to right, the initial scalar field value $\phi_0$ increases while the coupling $\epsilon$ decreases. Consequently, according to Eq.~\eqref{eq. oscillation redshift}, the oscillation redshift $z_{\rm osc}$ decreases, implying that the scalar field begins to oscillate at a later epoch. As a result, the extra energy density $\rho_{\rm extra}$ contributes more to the total energy density at later times, especially close to matter-radiation equality (MRE), leading to a larger expansion rate prior to the recombination. This reduces the sound horizon and, through
Eq.~\eqref{eq. H0}, yields a larger inferred value of the present-day Hubble parameter $H_0$. For every point in the $(\epsilon,\phi_0)$ parameter space, we compute the corresponding value of $H_0$ using Eq.~\eqref{eq. H0}. The orange
and cyan regions in Fig.~\ref{fig:scalarparspace} denote the parameter space for which the predicted $H_0$ lies within the SH0ES $1\sigma$
($72.26 \leq H_0 \leq 74.34~{\rm km\,s^{-1}\,Mpc^{-1}}$) and $2\sigma$
($71.22 \leq H_0 \leq 75.38~{\rm km\,s^{-1}\,Mpc^{-1}}$) intervals,
respectively, corresponding to the SH0ES determination
$H_0 = 73.30 \pm 1.04~{\rm km\,s^{-1}\,Mpc^{-1}}$.

Using the publicly available package \texttt{alterbbn} \cite{Arbey:2011zz, Arbey:2018b}, we determine that the maximum allowed extra DM energy density at the time of BBN ($T = 1 $ MeV) is  $10^{-2}$ times the photon density at BBN. DM densities larger than this value are excluded at the $2\sigma$ (95\%\,\text{C.L.}) level by the BBN constraints from the primordial helium mass fraction $Y_p$ and the deuterium-to-hydrogen ratio $\left( \frac{\text{D}}{\text{H}} \right)_p$. The values of $Y_p$ and $\left( \frac{\text{D}}{\text{H}} \right)_p$  are taken to be $0.245 \pm 0.003$ \cite{NavasPDG2024, Aver:2021qhb, Valerdi:2019qao, Fernandez:2019stz1433, Kurichin:2021stab215, Hsyu:2020phlek, Valerdi:2021stab1543} and $\left(25.47 \pm 0.29\right)\times 10^{-6}$ \cite{NavasPDG2024, Cooke:2013cba, Cooke:2016rqh, Riemer-Sorensen:2014yda, Balashev:2015xga, Riemer-Sorensen:2017qea, ZavaryginEtAl2018, Cooke:2017cwo}, respectively.
In the $\Lambda\text{CDM}$ scenario, the ratio of DM density to photon density at BBN ($T = 1 $ MeV) is of the order of $10^{-6}$. From Eq.\eqref{eq. evolution of rho4}, we see that in our scenario, at BBN, the sterile neutrino DM density is  $m^{\text{max}}_{\nu_4, \text{eff}} / m_{\nu_4}$ times larger than its value in the $\Lambda\text{CDM}$ scenario. So, in our scenario, this sets a maximum allowed value of $m^{\text{max}}_{\nu_4, \text{eff}} / m_{\nu_4}$ at BBN to be $ 8.8 \times 10^{3}$. We represent this bound by the hatched region in Fig.~\ref{fig:scalarparspace}. One can evade this BBN bound by decaying the scalar field before BBN by increasing the coupling $\epsilon$ such that $z_{\text{osc}}$ becomes greater than the redshift of BBN. In Fig.~\ref{fig:scalarparspace}, in the white region at the top left corner, the decay of $\phi$ happens before BBN. But for these values of $\epsilon$ and $\phi_0$, the \textit{Hubble tension} cannot be alleviated for any $m_{\nu_4}$. 
\begin{figure} 
   \centering
        \includegraphics[width=0.9\linewidth]{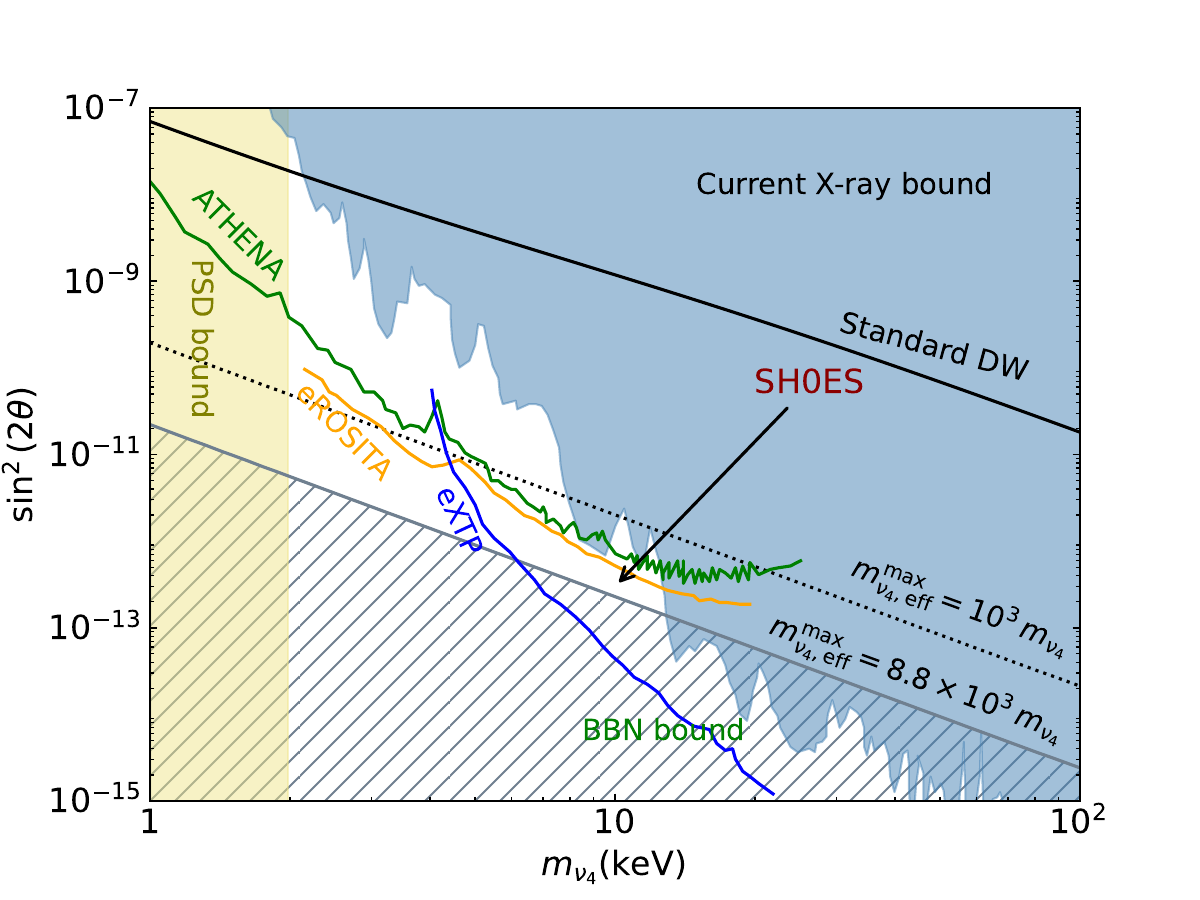}
        \caption{\small
         Sterile neutrino parameter space in presence of our new interaction term. In the white region DM relic and Hubble tension both can be satisfied. The black dotted and gray solid lines correspond to relic line for $m^{\text{max}}_{\nu_4 , \text{eff}} / m_{\nu_4} = 10^3$ and $8.8 \times 10^3$, respectively.
         X-ray and  phase-space density (PSD) bounds are denoted by blue and light yellow region respectively. The sensitivity curves of future X-ray missions \cite{Dev:2025} are shown in green (ATHENA), in blue (eXTP), and in orange (eROSITA). The black solid line is the standard Dodelson-Widrow line.  
        }
    \label{fig:parameter space}
\end{figure}

We depict the resulting sterile neutrino parameter space in $ \sin^2{2\theta} - m_{\nu_4}$ plane in Fig.~\ref{fig:parameter space}.  The solid black line is the Dodelson-Widrow line corresponding to the absence of new non-standard interactions. In the presence of a new interaction given in Eq.\eqref{eq. Lagrangian} viable parameter space is opened up without violating current X-ray bounds from Chandra \cite{Hofmann:2016urz}, XMM-Newton \cite{Borriello:2011un}, NuSTAR \cite{Krivonos:2024yvm}, and XRISM \cite{Yin:2025xad} telescopes.  The X-ray constraint is marked by the blue region.
The black dotted line and gray solid line correspond to the relic lines for $m^{\text{max}}_{\nu_4 , \text{eff}} / m_{\nu_4} = 10^3$ and $8.8 \times 10^3$, respectively. As noted earlier, the more we increase the $m^{\text{max}}_{\nu_4, \text{eff}} / m_{\nu_4}$ ratio, the less mixing we need between the active and sterile neutrino sector. The hatched region in Fig.~\ref{fig:parameter space} is excluded by BBN.
The white region of Fig.~\ref{fig:parameter space} corresponds to the parameter space in which the modified background expansion history yields the present-day value of the Hubble parameter, $H_0$, as measured by the SH0ES collaboration. The light-yellow region is excluded by the phase-space density constraint applicable to fermionic dark matter candidates, derived from observations of dwarf galaxies~\cite{Tremaine:1979we, Boyarsky:2008ju, Bezrukov:2024qwr}.
Following reference~\cite{Dev:2025}, we also add future sensitivities of upcoming missions of ATHENA (green line) \cite{Barret:2019qaw}, eROSITA (orange line) \cite{eROSITA:2012lfj}, and eXTP (blue line) \cite{Malyshev:2020hcc}. The parameter region of interest lies in the sweet spot of the sensitivities of future missions.

\section{Conclusion}
We have studied the production of sterile neutrino dark matter before BBN via the \textit{Dodelson-Widrow} (DW) mechanism in the presence of a new interaction term given in Eq.\eqref{eq. Lagrangian}. This interaction universally couples sterile and active neutrinos to a real homogeneous scalar field ($\phi$) quadratically and generates the observed relic. The scalar field gives an effective mass ($m_{\nu_4 ,\, \text{eff}}$) to the sterile neutrino. Due to this enhanced effective mass, production of the sterile neutrino from the active neutrino happens at high temperature. This results in a smaller vacuum mixing angle ($\sin^22\theta$)  needed for the DM relic generation compared to the standard DW scenario. We have used this feature to allow a smaller  $\sin^22\theta$ by increasing $m^{\text{max}}_{\nu_4,\, \text{eff}}$ in the parameter space of the sterile neutrino. In the standard DW scenario, this region of smaller $\sin^22\theta$ yields an under-abundant DM relic.

Besides that, we have demonstrated that the modified background cosmological evolution in this region of the sterile neutrino parameter space can yield the present-day value of the Hubble parameter, $H_0$, consistent with the SH0ES measurement for suitable choices of the coupling $\epsilon$ and the initial scalar-field value $\phi_0$. This enhancement arises because the sterile neutrino dark matter contributes an energy density larger than that in the standard $\Lambda$CDM cosmology by a factor of $m_{\nu_{4,\, \text{eff}}}/m_{\nu_4}$ during the early universe. Consequently, the Hubble expansion rate prior to the recombination epoch increases, reducing the sound horizon at photon decoupling and leading to a larger inferred present-day Hubble parameter.

Moreover, an increased Hubble rate during BBN leads to a larger neutron-to-proton ratio at freeze-out, thereby enhancing the primordial helium and deuterium abundances beyond the observed limits \cite{NavasPDG2024}. The corresponding constraint on $\sin^2 2\theta$ translates into a relic-density contour defined by $m^{\text{max}}_{\nu_4,\,\text{eff}} / m_{\nu_4} = 8.8 \times 10^3$. After imposing the astrophysical constraints, the phase-space density bound for fermionic dark matter, and the BBN constraint, we identify a parameter region where sterile neutrino masses in the range $2$--$13~{\rm keV}$ and vacuum mixing angle $\sin^22\theta \sim 10^{-8}$--$10^{-13}$,  yield the background value of $H_0$ compatible with the SH0ES measurement. Interestingly, this entire region will be probed by upcoming X-ray missions such as ATHENA \cite{Barret:2019qaw}, eROSITA \cite{eROSITA:2012lfj}, and eXTP \cite{Malyshev:2020hcc}.

We emphasize, however, that the present analysis is restricted to the background cosmological evolution. A quantitative assessment of whether the model alleviates the Hubble tension requires a rigorous cosmological likelihood analysis that includes CMB anisotropies, BAO, Type Ia supernovae, and large-scale structure data. Such an analysis necessitates implementing the model in a Boltzmann equation solver, such as CLASS \cite{DiegoBlas:2011}, or CAMB \cite{Lewis:1999bs}, and is left for future work.

\vspace{\baselineskip}

\acknowledgments
We thank Srubabati Goswami and Navaneeth Poonthottathil for helpful discussions. We thank the anonymous referee for their insightful comments on our manuscript.
D.C. acknowledges funding from the Indian Space Research Organisation (ISRO) under grant STC/PHY/2024427Q. M.S.I. would like to thank the MHRD, Government of India for the research fellowship. 
D.C. also acknowledges support from an initiation grant IITK/PHY/2019413 at IIT Kanpur and funding from the ANRF, Government of India, under grant ANRF/CRG/2021/007579.


\bibliographystyle{apsrev4-2}
\bibliography{main}

\end{document}